%% file: paper.tex
\documentclass[twocolumn,showpacs,aps,prl,superscriptaddress]{revtex4}

\usepackage{xspace}
\usepackage{graphicx}
\usepackage{dcolumn}
\usepackage{amsmath}
\usepackage{epsfig}

\input{babarsym}
\input{symbols}

 \def\er #1 #2 { $#1 \pm #2$ }
 \def\bra #1 #2 #3 #4 { $#1 ^{+#2} _{-#3} \pm #4 $ }

 \newcommand {\mhad}{\mbox{$M_{Xs}$}}

\newcommand{\BABARPubYear}     {04}
\newcommand{\BABARPubNumber}  {012}

\newcommand{\SLACPubNumber} {10386}

\def\figurebox#1#2#3{%
    \def\arg{#3}%
    \ifx\arg\empty
    {\hfill\vbox{\hsize#2\hrule\hbox to #2{\vrule\hfill\vbox to #1{\hsize#2\vfill}\vrule}\hrule}\hfill}%
    \else
    {\hfill\epsfbox{#3}\hfill}%
    \fi}

\begin{document}
\preprint{\babar-PUB-\BABARPubYear/\BABARPubNumber} 
\preprint{SLAC-PUB-\SLACPubNumber} 
%
%% Needed in final document
\begin{flushleft}
\babar-PUB-\BABARPubYear/\BABARPubNumber\\
SLAC-PUB-\SLACPubNumber\\
%hep-ex/\LANLNumber\\[10mm]
\end{flushleft}
\title{
{\large \bf \boldmath
Measurement of the Direct \CP\ Asymmetry in $b\to s \gamma$ Decays} 
}
%
% author list; 
\input authors_feb2004.tex
\date{\today}% It is always \today, today, but you may specify any date with \date.
\begin{abstract}
We describe a measurement of the direct \CP\ asymmetry between inclusive $\b\to s\gamma$ 
and $\bbar \to\sbar\gamma$ decays. This asymmetry is expected to be less than 0.01 in the Standard Model, 
but could be enhanced up to about 0.10 by new physics contributions. We use a sample of 89 million 
\BB\ pairs recorded with the \babar\ detector at \pep2, from which we reconstruct a set 
of 12 exclusive $b\to s\gamma$ final states containing one charged or neutral kaon and one to three pions. 
We measure an asymmetry of $A_{\CP}(\b\to s\gamma) = 0.025\pm 0.050(stat.)\pm 0.015(syst.)$, corresponding to an allowed range of 
$-0.06  < A_{\CP}(\b\to s\gamma)< +0.11$ at 90\%\ confidence level.  
\end{abstract}
\maketitle
%%
%% --------- Introduction ----------------
%%
The inclusive decay $\b\to s\gamma$ is a flavor--changing neutral current process described by a radiative penguin loop 
diagram. The world average branching fraction is $(3.5\pm 0.5)\times 10^{-4}$~\cite{PDG} 
in good agreement with recent theoretical predictions~\cite{TheoryBF}. 
Earlier experimental values of the branching fraction have been used to constrain new physics beyond the 
Standard Model~\cite{NewPhysics}.
A measurement of the direct \CP\ asymmetry between $\b\to s\gamma$ and $\bbar\to \sbar\gamma$ decays provides an 
independent and significant test of these predictions. 
In the Standard Model the dominant loop contribution contains a top quark, 
with other contributions being suppressed by CKM factors and the GIM mechanism. 
The lack of interference 
between comparable amplitude contributions leads to a rather small predicted asymmetry~\cite{Hurth}:
\begin{equation}
A^{TH}_{\CP} = \frac{\Gamma{(\b\to s\gamma)}-\Gamma{(\bbar\to \sbar\gamma})}{\Gamma({\b\to s\gamma)}+\Gamma{(\bbar\to \sbar\gamma})} = 0.0044^{+0.0024}_{-0.0014}
\end{equation} 
which has little sensitivity to the photon energy cut--off or to the distribution of hadronic final states.
The dominant errors are due to the uncertainty of the charm quark mass and the choice of the perturbative scale.
The inclusion of contributions to the loop beyond the Standard Model can increase the predicted asymmetry up to about 0.10~\cite{Hurth}. 

There is a previous measurement of direct \CP\ asymmetry~\cite{CLEO}
in a sum of $\b\to s\gamma$ and $\b\to \d\gamma$ decays.
In the Standard Model, the total of the $\b\to s\gamma$ and $\b\to \d\gamma$ asymmetries
is exactly zero in the U-spin symmetry limit, $m_d = m_s$, as a consequence of CKM unitarity~\cite{Soares}. 
The measurement in Ref.~\cite{CLEO} gives
$-0.27<0.965 \times A_{\CP}(\b\to s\gamma) + 0.02 \times A_{CP}(b\to d\gamma)<0.10$.

%%
%% --------- Detector ----------------
%%
We use a sample of $(88.9\pm 1.0)\times 10^6$ \BB\ pairs collected 
at the \FourS\ resonance with the \babar\ detector at the \pep2\ asymmetric $e^+e^-$ collider.
A detailed description of the detector can be found elsewhere~\cite{detector}. 
For this analysis the 
most important detector elements are the forty--layer drift chamber, situated in a 1.5\,T solenoidal
magnetic field, which measures charged particle momenta,
the CsI(Tl) electromagnetic calorimeter, which measures the 
energies of the photons, and the detector of internally reflected Cherenkov light (DIRC), 
which is used to identify charged kaons.  

We reconstruct $\b\to s\gamma$ decays as the sum of twelve exclusive final states: 
\begin{flushleft}
$\Bub\to \Km\piz\gamma, \Km\pip\pim\gamma, \Km\piz\piz\gamma, \Km\pip\pim\piz\gamma$ \\
$\Bzb\to \Km\pip\gamma, \Km\pip\piz\gamma, \Km\pip\piz\piz\gamma, \Km\pip\pim\pip\gamma$ \\
$\Bub\to \KS\pim\gamma, \KS\pim\piz\gamma, \KS\pim\piz\piz\gamma, \KS\pim\pip\pim\gamma$ \\
\end{flushleft}
and measure the yield asymmetry with respect to their charge conjugate decays $\bbar\to \sbar\gamma$. 
The identification of charged kaons removes $\b \to  d \gamma$ decays.
We do not use $\Bzb$ decays to final states with $\KS$ to determine the direct \CP\ asymmetry,
since these are not flavor--specific, but we study them to understand systematic effects.

The high energy photon is detected from an isolated energy cluster in the calorimeter, with
shape consistent with a single photon, and energy $E_{\gamma}^*>1.8$~\gev\ in the $e^+e^-$ center--of--mass frame. 
A veto is applied to the high energy photons that combined with another photon form either a $\piz$ within
the mass range 117--150~\mevcc\ or an $\eta$ within the mass range 524--566~\mevcc. 

Neutral kaons are reconstructed as $\KS\to\pip\pim$ candidates with an invariant mass 
within 9~\mevcc\ of the nominal mass~\cite{PDG}, and a transverse flight distance $>2$~mm from the 
primary event vertex. Charged kaons are tracks identified as kaons from information in the DIRC. 
The remaining tracks are considered to be charged pions.
Both charged and neutral kaons are required to have a laboratory momentum 
$>$ 0.7~\gevc. Above this threshold the rate for charged pions to be mis-identified as kaons is $<$ 2.0\%. 

Neutral pions are reconstructed from pairs of photons with energies $>$ 30~\mev.  
A $\piz$ mass cut is applied between 117 and 150~\mevcc. 
Charged and neutral pions are required to have laboratory momenta 
$>$ 0.5, 0.3 or 0.2~\gevc\ for states with 1, 2 or 3 pions, respectively, to reject combinatoric background.

The mass of the hadronic system, $X_{s}$, formed from the kaon and pions is required to be between 
0.6~\gevcc\ and 2.3~\gevcc, corresponding to a photon energy threshold $E_{\gamma} > 2.14$~\gev\ in the \B\ rest frame.   

The signal Monte Carlo sample is generated according to Ref.~\cite{KaganNeubert}, 
which predicts that ($83 \pm 5$)\% of the $b\to s\gamma$ spectrum is above our photon energy threshold.
We use JETSET~\cite{JETSET} to hadronize the system of the strange and spectator quarks.
Within the selected hadronic mass range, the twelve final states constitute 48\%\ of the total rate.
If we also include the $\Bzb$ decays to $\KS$ and equate the decays to \KL\
with those to \KS, this increases to 73\%\ of the total rate.
As a part of our analysis, we check the dependence of the asymmetry on 
the hadronic mass and final state.

%%
%% --------- Analysis ----------------
%%

Most of the background in this analysis arises from continuum production of a high 
energy photon, either by initial state radiation, or from the decays of 
$\piz$ and $\eta$ mesons. We remove 86\% of these backgrounds by selections on 
the angle between the thrust axis of the \B\ meson candidate and the thrust axis of all
the other particles of the event, $|\cos\theta_T^*| < 0.80$, and the angle between the \B\ candidate 
and the beam axis, $|\cos\theta_B^*|<0.80$, both defined in the $e^+e^-$ 
center--of--mass system. We then use a neural network to combine information from a 
set of event shape variables, including a set of energy flow cones. 
This halves the continuum background compared to our 
initial selection.

In 12\%\ of the signal events, we can identify an electron or muon from the decay of the other \B~\cite{btagging}. 
This is a very effective signature for removing continuum background, so the 
remaining background in this sample comes mostly from other \B\ decays. 
We present separately our results for the sample of events which are lepton--tagged.

Exclusive $\b\to s\gamma$ decays are characterized by two kinematic variables:
the beam--energy substituted mass, $m_{ES}= \sqrt{(\sqrt{s}/2)^2- p^{*2}_\B}$, 
and the energy difference between the \B\ candidate and the beam energy, 
$\Delta E = E_\B^* - (\sqrt{s}/2)$, where $E_\B^*$ and $p_\B^*$ are the energy and 
momentum of the \B\ candidate in the $e^+e^-$ center--of--mass frame, and 
$\sqrt{s}$ is the total center--of--mass energy. 
We require candidates to have $|\Delta E|<0.10$~\gev, and 
remove multiple candidates in each event by selecting the one
with the smallest value of $|\Delta E|$.
This technique is $>$ 90\%\ efficient when the true $\b \to s \gamma$
decay is among the reconstructed candidates.
We then fit the $m_{ES}$ distribution between 5.22 and 5.29~\gevcc\ to extract the signal yield. 
When calculating $m_{ES}$, the value of $p^*_\B$ is corrected for the tail of the 
high energy photon response function by scaling the measured $E_{\gamma}^*$ to the value that would 
give $\Delta E = 0$, the value expected for true signal.

\begin{figure}
\resizebox{\columnwidth}{!}{%
\includegraphics{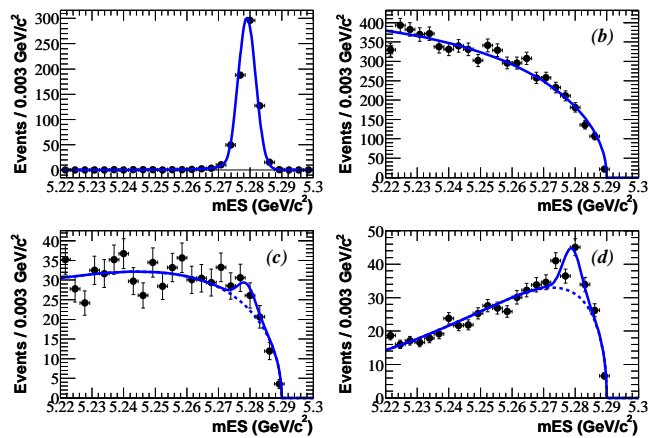}}
\caption{
Monte Carlo simulations of the four contributions to the beam--energy substituted mass distribution 
of events selected as $b\to s\gamma$, with the corresponding fits: 
(a) signal, (b) continuum, (c) \BB\ decays and (d) cross--feed.
The plots are normalized to the luminosity of our data sample.}
\label{fig1}
\end{figure}

In order to fit the \mes\ distribution in data, we need to understand the different components of 
the signal and background events. 
We have identified the following four contributions as shown in 
Figure~\ref{fig1}. The signal events are described by a Crystal Ball 
function~\cite{CBfunction} with a resolution $\sigma(m_{ES}) = 2.2$~\mevcc. 
The continuum background is described by an ARGUS shape~\cite{ARGUS}, 
which is cross--checked by a fit to a sample of 9.6$fb^{-1}$ of data taken 40~\mevcc\ below the 
\FourS\ resonance. We use a \BB\ Monte Carlo sample to model 
the background from \B\ decays other than $\b\to s\gamma$, 
which is significant for $X_{s}$ masses above 1.9~\gevcc. This background is described by 
the sum of an ARGUS shape and a peaking component which is modelled by the signal shape. 

The last background component is cross--feed from incorrectly 
reconstructed $\b\to s\gamma$ events. This is modelled by the signal Monte Carlo sample, where 
we identify events reconstructed in the 
wrong final state. Cross--feed occurs when 
the true $\b \to s \gamma$ decay is not among the reconstructed candidates, 
or in a multiple candidate event when the wrong candidate is chosen. 
The shape of the cross--feed is described by the sum of an ARGUS shape 
and a peaking signal shape. 
We regard cross--feed as a background to be subtracted.

\begin{figure}
\resizebox{\columnwidth}{!}{%
\includegraphics{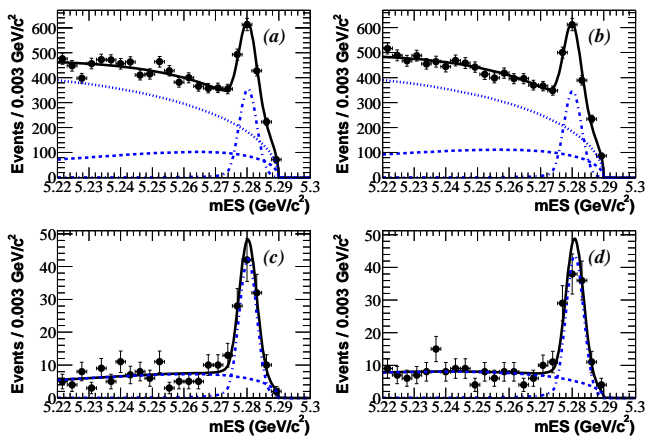}}
\caption{Fits to the beam--energy substituted mass distributions in data events for:
(a) all $\b\to s\gamma$, (b) all $\bbar\to\sbar\gamma$, 
(c) lepton--tagged $\b\to s\gamma$ and (d) lepton--tagged $\bbar\to\sbar\gamma$ decays.
Contributions are shown from peaking Crystal Ball (dotted--dashed), fixed continuum ARGUS shape (dotted) 
and free \BB\ and cross--feed ARGUS shape (dashed).}
\label{fig2}
\end{figure}

We fit the data \mes\ distributions separately for each flavor. 
For the total sample, the fit function is parametrized by two ARGUS shapes
and a Crystal Ball function.
One ARGUS shape is fixed to be as the continuum ARGUS shape, 
while the other one is free to represent the sum of the non-peaking \BB\ and cross--feed backgrounds.   
The Crystal Ball function fits the combination of the peaking components. 
For the lepton--tagged sample, we use only one free ARGUS shape and a Crystal Ball function.
In all cases we use an unbinned maximum likelihood fit.
The fitting technique has been validated with a large sample of Monte Carlo simulated events.
In Figure~\ref{fig2} we present the final fits to the $m_{ES}$ distributions
for $\b\to s\gamma$ and $\bbar\to\sbar\gamma$ events. The lower plots 
are for the lepton--tagged sample. All the fits have $\chi^2$ per degree--of--freedom close to 1,
if we make a fit to a binned distribution as shown in Figure~\ref{fig2}.
The sum of events in the \b\ and \bbar\ peaks is $1644\pm 72$, of which $201\pm 18$ are 
lepton--tagged.
To get the true signal yields these have to be corrected for the predicted 
yield of peaking \BB\ and cross--feed backgrounds from Monte Carlo samples (see Figure~\ref{fig1}), 
which is $88\pm 27$, where $10\pm 8$ are lepton--tagged.

The direct \CP\ asymmetry is calculated from:
\begin{equation}
A_{\CP} = \frac{1}{\langle D\rangle}\left(\frac{(n-\bar{n})}{(n+\bar{n})}-\frac{\Delta D}{2}\right)-A_{\CP}^{DET}
\end{equation} 
where $n$ and $\bar{n}$ are the numbers of observed $\b\to s\gamma$ and $\bbar\to\sbar\gamma$ events
after the peaking background is subtracted, 
$\Delta D = 2 (\bar{w} -w)$ is the difference in the wrong flavor--fraction between \b\ and \bbar\ decays,
and $\langle D\rangle = 1 - (w+\bar{w})$ is the dilution factor from the average wrong flavor--fraction.
$A_{\CP}^{DET}$ is the flavor--asymmetry of the detector.
We find $\Delta D = 0.001 \pm 0.002$ and $\langle D\rangle =0.989 \pm 0.001$ from Monte Carlo samples.
The small wrong--flavor fraction is due to charged pions mis-identified as charged kaons. 

We need to correct the measured value of $A_{\CP}$ for the flavor--asymmetry of the detector $A_{\CP}^{DET}$. 
While it is known that 
the kaon--nucleon cross--sections are asymmetric at low momenta, there are few accurate measurements~\cite{PDG}. 
This means that our Monte Carlo sample is not expected to model correctly the asymmetries due to the interactions of kaons
with the inner part of the detector. 
The kaon identification efficiency of the DIRC for reconstructed tracks is measured with a control sample of 
kaons from \Dstar\ decays. Averaging over the kaon spectrum in $b\to s\gamma$ events we obtain a 
small asymmetry of $-0.002 \pm 0.001$ from particle identification.
We measure the overall detector asymmetry of the data events in our $m_{ES}$ and $\Delta E$ sidebands, increasing 
the statistics by removing the neural network cut. Most of these events are from the continuum, where 
we do not expect any physics mechanism to generate a flavor--asymmetry. We observe a significant asymmetry 
for kaon momenta below 1~\gevc. The asymmetry as a function of the kaon momentum is applied
to the signal Monte Carlo to determine what shift should be applied to the data. 
This gives an overall flavor--asymmetry correction $A_{\CP}^{DET}=-0.014\pm 0.015$.  

Table~\ref{results} presents the measured signal yields and corrected \CP\ asymmetries.
The lepton--tagged results are consistent with the results for the total sample.
We divide the total sample into four bins in $X_{s}$ mass, and observe no significant 
mass dependence of the asymmetry. The first bin corresponds to the $\Kstar(892)$ resonance, 
for which the world average asymmetry from studies of exclusive $\B\to \Kstar\gamma$ decays is 
$A_{\CP} (\B\to \Kstar\gamma) = -0.005\pm 0.037$~\cite{PDG}. Our result is consistent with this average.

We divide our total sample into three types of decay mode: $\Bz(\Bzb)\to \Kpm$, 
$\Bpm\to \Kpm$ and $\Bpm\to \KS$. We observe a discrepancy of 
2.3$\sigma$ between the two \Bpm\ categories which we regard as a 
statistical fluctuation, since it is not correlated with a specific final state or 
hadronic mass bin. The combination of the \Bpm\ samples 
is consistent with a null asymmetry, as is the \Bz\ sample.
 
\begin{table}[htbp]
\caption{Signal yields and \CP\ asymmetries for total and only lepton--tagged event
samples. 
The total sample is also divided up into four bins in $X_{s}$ mass in~\gevcc, and 
into three types of decay modes. The errors on $n$ and $\bar{n}$ are statistical only, 
while for $A_{\CP}$ we quote the additional systematic error from the detector asymmetry.}
\begin{center}
\begin{tabular}{|l|c|c|c|}  
\hline
Sample  & $n$  & $\bar{n}$ & $A_{\CP}$ \\
\hline
Total Sample & 787 $\pm$ 54 & 769 $\pm$ 54 & 0.025 $\pm$ 0.050 $\pm$ 0.015 \\
\hline
Lepton--tagged & 91 $\pm$ 14 & 100 $\pm$ 13 & -0.04 $\pm$ 0.10 $\pm$ 0.02 \\
\hline
\mhad=0.6-1.1 & 378 $\pm$ 32 & 396 $\pm$ 33 & 0.003 $\pm$ 0.059 $\pm$ 0.015 \\
\mhad=1.1-1.5 & 162 $\pm$ 22 & 136 $\pm$ 23 & 0.11 $\pm$ 0.11 $\pm$ 0.02 \\
\mhad=1.5-1.9 & 139 $\pm$ 19 & 124 $\pm$ 21 & 0.07 $\pm$ 0.11 $\pm$ 0.03 \\
\mhad=1.9-2.3 & 101 $\pm$ 29 & 67 $\pm$ 36 & 0.23 $\pm$ 0.30 $\pm$ 0.04 \\
\hline
\Bz              & 455 $\pm$ 36 & 447 $\pm$ 38 & 0.015 $\pm$ 0.059 $\pm$ 0.014 \\
$\Bpm \to \Kpm$  & 229 $\pm$ 31 & 148 $\pm$ 30 & 0.22 $\pm$ 0.12 $\pm$ 0.02 \\
$\Bpm \to \KS$   & 100 $\pm$ 24 & 166 $\pm$ 25 &-0.20 $\pm$ 0.14 $\pm$ 0.03 \\
\hline
\end{tabular}
\end{center}
\label{results}
\end{table} 

The dominant systematic error in our measurement is the uncertainty 
of 0.015 in the flavor--asymmetry of the detector.
For the lepton--tagged sample we add an additional systematic uncertainty 
of 0.010 to account for a possible charge asymmetry in the lepton tagging efficiency.
This is derived from studies of control samples~\cite{btagging}. 
 
We have tested the effect of possible flavor asymmetries in the
peaking cross--feed and \BB\ backgrounds by varying them within the current experimental bounds (90\% C.L.). 
We added a 0.10 asymmetry to the cross--feed events, and a 0.02 asymmetry to the 
peaking background from \BB\ decays, which comes primarily
from $\B \to D^{(*)}\rho$ decays. The change in our measured asymmetry due to these changes in
the cross--feed and \BB\ flavor--asymmetries is 0.004, which gives a negligible contribution
to the error.

We have checked that the parameters of the ARGUS shapes and Crystal Ball functions 
are the same for both flavors within 1$\sigma$, so the detector asymmetry 
is simply an overall normalization difference between the two samples. 
We have also checked that the neural net distributions for signal and continuum 
background are flavor--symmetric. 

Our estimates of the cross-feed 
background and the detector asymmetry correction, $A_{\CP}^{DET}$, depend on the mix 
of final states in our signal Monte Carlo sample. We check these, also using
information from $\Bzb$ decays to final states with $\KS$, by varying the ratios of 
final states with \Kp\ or \KS, and $\piz$ to $\pip$ measured in our data by $\pm 3 \sigma$. 
Note that the measured ratios are consistent with our signal Monte Carlo. 
Changing the ratios has no significant effect on the cross--feed or the detector asymmetry correction. 

Our final result for the direct \CP\ asymmetry in $\b\to s\gamma$ is 
$A_{\CP}= 0.025\pm 0.050\pm 0.015$ for the total sample, and 
$A_{\CP} = -0.04\pm 0.10\pm 0.02$ for the lepton--tagged sample. 
The total sample provides the best constraint, $-0.06 < A_{\CP}< +0.11$ at 90\%\ confidence level.

% Input the pubboard acknowledgements file
\input acknow_PRL.tex

\end{document}

%% file: symbols.tex
\def\Dp      {\ensuremath{D^+}}
\def\Dm      {\ensuremath{D^-}}
\def\Dstar   {\ensuremath{D^{*}}}
\def\Dstarp  {\ensuremath{D^{*+}}}
\def\Dstarm  {\ensuremath{D^{*-}}}

\newcommand{\etal}{{\it et al.}}

\def\ggstar{\ensuremath{\gamma\gamma^*}}

\def\bbar  {\ensuremath{\overline b}}
\def\piz   {\ensuremath{\pi^0}}

\def\ppz   {\ensuremath{\pi^0\pi^0}}
\def\KS    {\ensuremath{K^0_{\scriptscriptstyle S}}} %<<<
\def\KL    {\ensuremath{K^0_{\scriptscriptstyle L}}} %<<<

\def\BB    {\ensuremath{B\overline{B}{}}} %<<<
\def\BzBzb {\ensuremath{B^0 \overline{B}{}^0}} %<<<
\def\Bbar  {\ensuremath{\overline{B}{}}} %<<<new
\def\Dbar  {\ensuremath{\overline{B}{}}} %<<<new
\def\Kbar  {\ensuremath{\overline{B}{}}} %<<<new
\def\jpsi  {\ensuremath{{J\mskip -3mu/\mskip -2mu\psi\mskip 2mu}}} %<<<
\def\FourS {\ensuremath{\Upsilon{\rm( 4S)}}} %<<<
\def\Y#1S  {\ensuremath{\Upsilon\rm(#1S)}} %<<<new

\def\ep		{\ensuremath{e^+}}
\def\en		{\ensuremath{e^-}}		% electron negative (\em is taken)

\def\mumu	{\ensuremath{\mu^+\mu^-}}
\def\mup	{\ensuremath{\mu^+}}
\def\mun	{\ensuremath{\mu^-}}	% muon negative (\mum is taken)

\def\taup	{\ensuremath{\tau^+}}
\def\taum	{\ensuremath{\tau^-}}

\def\pip	{\ensuremath{\pi^+}}
\def\pim	{\ensuremath{\pi^-}}

\def\pipi       {\ensuremath{\pi^+\pi^-}}

\def\kaon	{\ensuremath{K}}

\def\BR		{\rm B}

\def\BF{$B$ Factory}

\def\kev  {\ensuremath{\mbox{\,ke\kern -0.08em V}}} %<<<
\def\mev  {\ensuremath{\mbox{\,Me\kern -0.08em V}}} %<<<
\def\gev  {\ensuremath{\mbox{\,Ge\kern -0.08em V}}} %<<<
\def\gevc {\ensuremath{\mbox{\,Ge\kern -0.08em V$\!/c$}}} %<<<
\def\mevc {\ensuremath{\mbox{\,Me\kern -0.08em V$\!/c$}}} %<<<
\def\gevcc{\ensuremath{\mbox{\,Ge\kern -0.08em V$\!/c^2$}}} %<<<
\def\mevcc{\ensuremath{\mbox{\,Me\kern -0.08em V$\!/c^2$}}} %<<<

\def\mus  {\ensuremath{\,\mu\mbox{s}}}

%% file: authors_feb2004.tex
% author list as of 02-Feb-2004 (597 authors)
%
\author{B.~Aubert}
\author{R.~Barate}
\author{D.~Boutigny}
\author{F.~Couderc}
\author{J.-M.~Gaillard}
\author{A.~Hicheur}
\author{Y.~Karyotakis}
\author{J.~P.~Lees}
\author{V.~Tisserand}
\author{A.~Zghiche}
\affiliation{Laboratoire de Physique des Particules, F-74941 Annecy-le-Vieux, France }
% ...
\collaboration{The \babar\ Collaboration}
\noaffiliation

%% file: acknow_PRL.tex
We are grateful for the excellent luminosity and machine conditions
provided by our \pep2\ colleagues, 
and for the substantial dedicated effort from
the computing organizations that support \babar.
The collaborating institutions wish to thank 
SLAC for its support and kind hospitality. 
This work is supported by
DOE
and NSF (USA),
NSERC (Canada),
IHEP (China),
CEA and
CNRS-IN2P3
(France),
BMBF and DFG
(Germany),
INFN (Italy),
FOM (The Netherlands),
NFR (Norway),
MIST (Russia), and
PPARC (United Kingdom). 
Individuals have received support from the 
A.~P.~Sloan Foundation, 
Research Corporation,
and Alexander von Humboldt Foundation.